\documentclass[final,3p,times]{elsarticle}
\usepackage[utf8]{inputenc}
\usepackage{graphicx}
\usepackage{bm}%
\usepackage{siunitx}
\usepackage{textcomp}
\usepackage{lineno}
\usepackage{xcolor}
\usepackage{subfigure}
\usepackage{amsmath}
\usepackage{textcomp}
\usepackage{multicol}
\usepackage{multirow}
\usepackage[figuresright]{rotating}
\usepackage{hyperref}
\biboptions{sort&compress}
\journal{NIMA}

\begin{document}

\begin{frontmatter}


\title{Modelling the Response of CLLBC(Ce) and TLYC(Ce) SiPM-Based Radiation Detectors in Mixed Radiation Fields with Geant4}



\author[a,b]{Jeremy~M.~C.~Brown}
\author[a]{Lachlan~Chartier}
\author[a]{David~Boardman}
\author[a]{John~Barnes}
\author[a]{Alison~Flynn}

\address[a]{Detection and Imaging, Australian Nuclear Science and Technology Organisation (ANSTO), NSW 2234, Australia}
\address[b]{Optical Sciences Centre, Department of Physics and Astronomy, Swinburne University of Technology, Hawthorn, VIC 3122, Australia}

\begin{abstract}
CLLBC(Ce) and TLYC(Ce) are novel scintillation materials capable of measuring mixed gamma ray and neutron radiation fields that have gained significant interest in the areas of space and nuclear safety/security science. To date Geant4, the world’s most popular Monte Carlo radiation modelling toolkit, has yet to be effectively used to simulate the full response of these materials when coupled to near ultra-violet Silicon PhotoMultipliers (SiPMs). In this work an experimentally validated Geant4 application has been developed with optimised material composition, optical data tables, and physics transport settings that is able to accurately simulate the response of CLLBC(Ce) and TLYC(Ce) SiPM-based radiation detectors under both gamma ray and neutron irradiation. Experimental benchmarking for five different radioactive sources (Co-60, Cs-137, Eu-152, Am-241, and Cf-252) illustrated that this developed Geant4 application was able to reproduce the position and structure of all major spectral features (full energy gamma ray photo-/neutron capture peaks, X-ray escape photopeaks, Compton edge, Compton backscatter peaks, and Compton plateau) to high level of accuracy. 

\end{abstract}

\begin{keyword}
CLLBC \sep TLYC \sep Geant4 \sep gamma/neutron detection \sep Geant4 optical physics
\end{keyword}

\end{frontmatter}


\section{Introduction}

Over the last 20 years several novel scintillation materials capable of measuring mixed gamma and neutron radiation fields have been developed targeting space and nuclear safety/security applications. Two of the more recent of these materials are the elpasolite crystals CLLBC(Ce) \cite{Shirwadkar2012,Hawrami2016a} and TLYC(Ce) \cite{Hawrami2015,Hawrami2016b}. CLLBC(Ce) and TLYC(Ce) have high (50000 photons per MeV \cite{Shirwadkar2012,Hawrami2016a}) and moderate (25000 photons per MeV \cite{Hawrami2015,Hawrami2016b}) light outputs respectively, and their optical emission spectra and decay times match well with the wavelength dependent photodetection efficiency and recharge times of standard near ultra-violet Silicon PhotoMultipliers (SiPMs). Both materials are also effectively transparent to their own optical emissions and can be grown to yield large volume crystals with high levels of uniformity. In head-to-head comparisons when bonded to SiPMs, CLLBC(Ce) and TLYC(Ce) have been shown to outperform the energy resolution of CsI(Tl) \cite{Hawrami2016a,Hawrami2016b} making them both excellent candidates for applications where size and mass need to be minimised (i.e. hand-held survey detectors, drones, etc.).  

CLLBC(Ce) and TLYC(Ce)’s ability to detect neutrons can primarily be attributed to the presence of Li-6 which has a high neutron capture cross-section \cite{Hawrami2016a,Hawrami2016b}:

\begin{equation}
\begin{aligned}
    \displaystyle {^6}Li + n \rightarrow {^4}He + {^3}H \qquad Q = 4.8 MeV\\
\end{aligned}
\label{eqn:1}
\end{equation}

Two primary methods are currently in use to differentiate between gamma ray or neutron detection events in both materials: 1) Pulse Shape Discrimination (PSD), and 2) Full Equivalent Energy Peak Discrimination (FEEPD) \cite{Hawrami2016a,Hawrami2016b}. PSD exploits the differences in light emission time structures that are generated from electron (gamma ray) and alpha particle (neutron) secondaries in these scintillator crystals due to their relative propagation times before reaching zero kinetic energy. Whilst this approach has been shown to work well, it requires extensive electronics to digitise and process the detector signal output on-the-fly. Whereas FEEPD exploits the fact that neutron capture by Li-6 in CLLBC(Ce) and TYLC(Ce) generate distinct peaks (3.2 and 1.8 MeV respectively) that reside at the upper edge of the relevant energy range for gamma rays in nuclear safety and security applications \cite{Hawrami2016a,Hawrami2016b}. The observed difference between the neutron capture full energy peaks of both materials with respect to the Q value of the reaction seen in Equation \ref{eqn:1} (4.8 MeV) is a direct result of crystal ionisation saturation and, in turn, the reduced light output from the secondary heavy particles known as the Birks’ effect \cite{Birks1951,Tretyak2010}.  

Of the available Monte Carlo radiation transport modelling toolkits, the Geant4 toolkit for the simulation of the passage of particles through matter \cite{G42003,G42006,G42016} is the most popular employed in the fields of space and nuclear safety/security. To date Geant4 has yet to be effectively used to simulate the full response of these materials when coupled to near ultra-violet SiPMs. This work outlines the development of an experimentally validated Geant4 version 10.7 application with optimal material composition, optical data tables, and physics transport settings capable of accurately simulating the response of Radiation Monitoring Devices Inc. (RMD) 1/2-inch CLLBC(Ce) and TLYC(Ce) SiPM-based radiation detectors. Of primary interest in this work is assessing Geant4’s ability to model FEEPD of neutron capture detection events for the two materials in mixed radiation fields. Within this Geant4 application electromagnetic, hadronic, and optical physics were implemented to simulate the transport of particles from the emission of the gamma ray/neutron down to the absorption of scintillation photons within each detector volume.  

\section{Method}
\subsection{Experimental Platform and Acquisition of Radiation Energy Spectra}

\begin{figure}[ht]   
   \centering
    \includegraphics[width=0.45\textwidth]{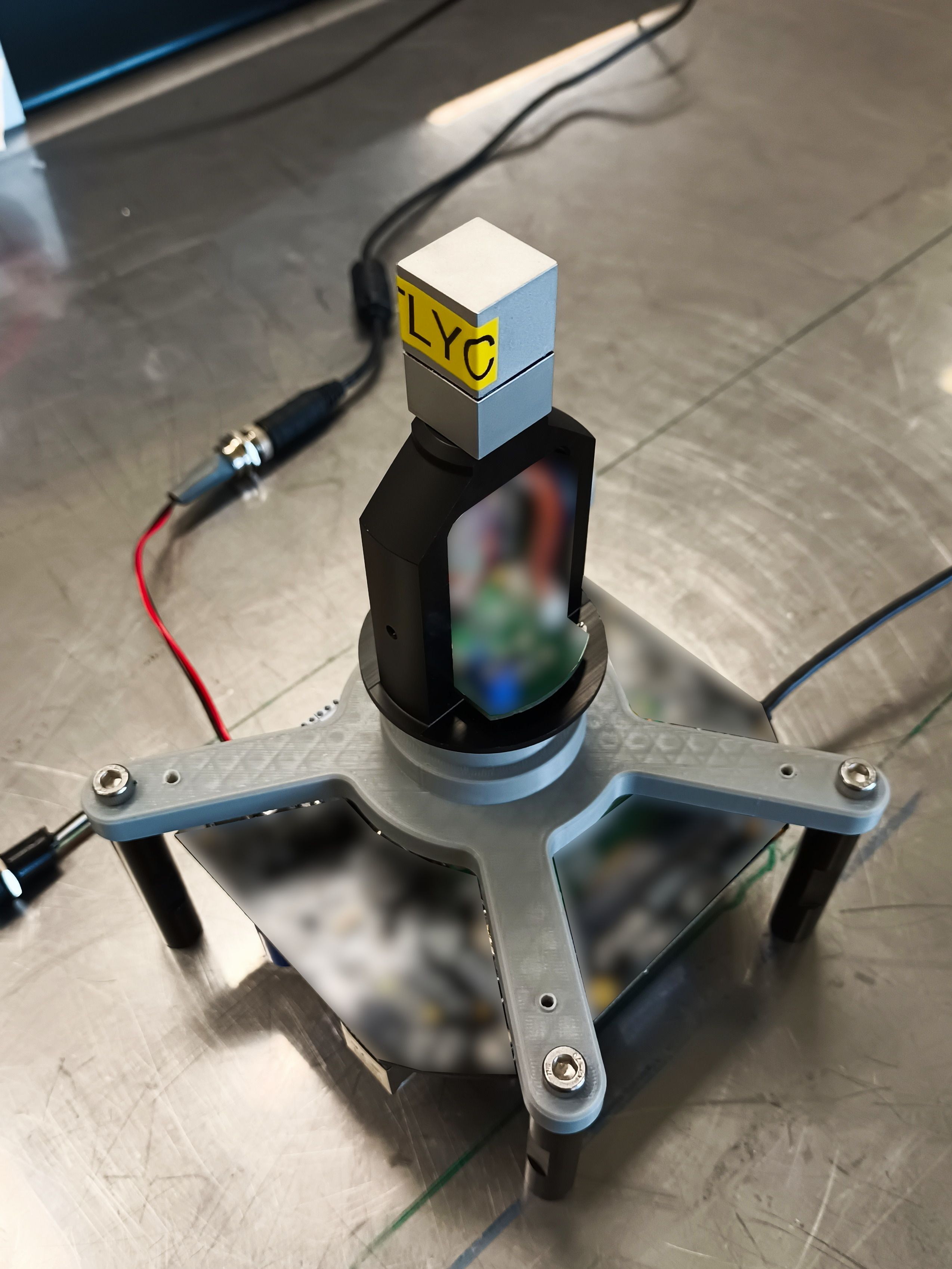}
\caption{RMD 1/2-inch CLLBC(Ce) and TLYC(Ce) SiPM-based radiation detector, signal processing electronics, and custom 3D printed plastic mounting platform that was developed at ANSTO. }
\label{fig:1}
\end{figure}

The RMD 1/2-inch CLLBC(Ce) and TLYC(Ce) SiPM-based radiation detector, signal processing electronics, and custom 3D printed plastic mounting platform that were utilised in the present study can be seen in Figure \ref{fig:1}. Each 1/2-inch RMD scintillator detector is composed of seven key components: aluminium housing, CLLBC(Ce) and TYLC(Ce) scintillator crystal, EJ-560 optical pad, Onsemi ARRAYJ-60035-4P-BGA SiPM, interface PCB, multilayer Teflon wrapping, and GORE diffuse reflector stack end caps. The outer Al housing is environmentally and optically isolated to protect both scintillator materials which are hygroscopic \cite{Hawrami2016a,Hawrami2016b}, and the combination of GORE diffuse reflector and multilayer Teflon wrapping was implemented to maximise scintillation light collection at the Onsemi SiPM. Each Onsemi SiPM is composed of a 2 $\times$ 2 array of 6 $\times$ 6 mm$^2$ pixels that each contain 22,292 35 $\times$ 35 \textmu m$^2$ Single Photon Avalanche Diodes (SPADs) \cite{Onsemi2021}. All pixel output signals are summed, and the total signal is fed into a XIA microDXP digital pulse processing unit to obtain integrated pulse energy spectra. The integration time for both the CLLBC(Ce) and TLYC(Ce) was set to 5 \textmu S which resulted in a dead time of less than 5\% for all measurements. 

Two different experimental configurations were used to measure the energy spectra of the RMD 1/2-inch CLLBC(Ce) and TLYC(Ce) SiPM-based radiation detectors under gamma ray and neutron irradiation with the detection platform that can be seen in Figure \ref{fig:1}. For the gamma ray measurements two different gamma ray emitting source types were used types: 1) three planar Isotrack QSA Global sources (Co-60, Cs-137, and Am-241) with 32 mm active diameters, and 2) a Spectrum Techniques encapsulated disk point source (Eu-152). Each source was placed directly on-top of the detector to maximise the signal with respect to background scatter off the surrounding environment. In the case of the neutron source measurements, a Cf-252 pellet was placed in the centre of a 195 mm diameter/225 mm tall Perspex cylinder and orientated approximately 100 mm away from each detector to moderate the neutron spectra towards a mean thermal energy. Individual background and energy spectrum measurements were undertaken for each radiation source-detector combination. 

\subsection{Geant4 Application Detector Geometry, Materials and Optical Data Tables}

\begin{figure}[ht]   
   \centering
    \includegraphics[width=0.45\textwidth]{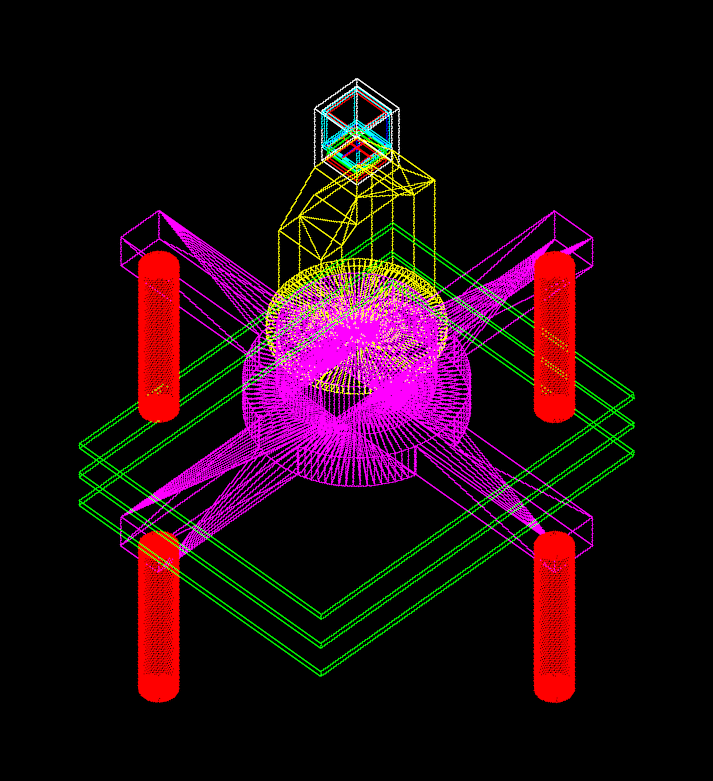}
\caption{A schematic of the implemented Geant4 geometry of the RMD 1/2-inch CLLBC(Ce) and TLYC(Ce) SiPM-based radiation detector, signal processing electronics, and custom 3D printed plastic mounting platform.}
\label{fig:2}
\end{figure}

A schematic of the implemented Geant4 geometry replicating the experimental set up outlined above can be seen in Figure \ref{fig:2}. Here the implemented simulation geometry can be separated into four primary structures: 1) laboratory benchtop, 2) custom 3D printed plastic mounting platform, 3) signal processing electronics, and 4) CLLBC(Ce) /TLYC(Ce) SiPM-based radiation detector. The first three of these structures were implemented as surrogate objects with simplified geometries. The laboratory benchtop was modelled as a two-layer table with a 2 mm thick stainless-steel top and 40 mm thick MDF bulk that spans the entire simulation geometry. The custom 3D printed plastic mounting platform was implemented via 3 sub-structures: 1) four 12 (D) $\times$ 60 (H) mm cylindrical aluminium legs, 2) a multi-layered 3D PLA printed 183 (D) $\times$ 183 (W) $\times$ 32 (H) mm cross-like mounting platform, and 3) a 5 mm thick 16 (D) $\times$ 48 (W) $\times$ 68.5 (H) mm O shaped acetal detector stick. Whereas the signal processing electronics were implemented as a three-layered stack of 102 (D) $\times$ 132 (W) $\times$ 2 (H) mm FR4 sheets spaced 10 mm apart.  

\begin{figure}[h]    
    \centering 
    \begin{subfigure}
    \centering 
        \includegraphics[width=0.45\textwidth]{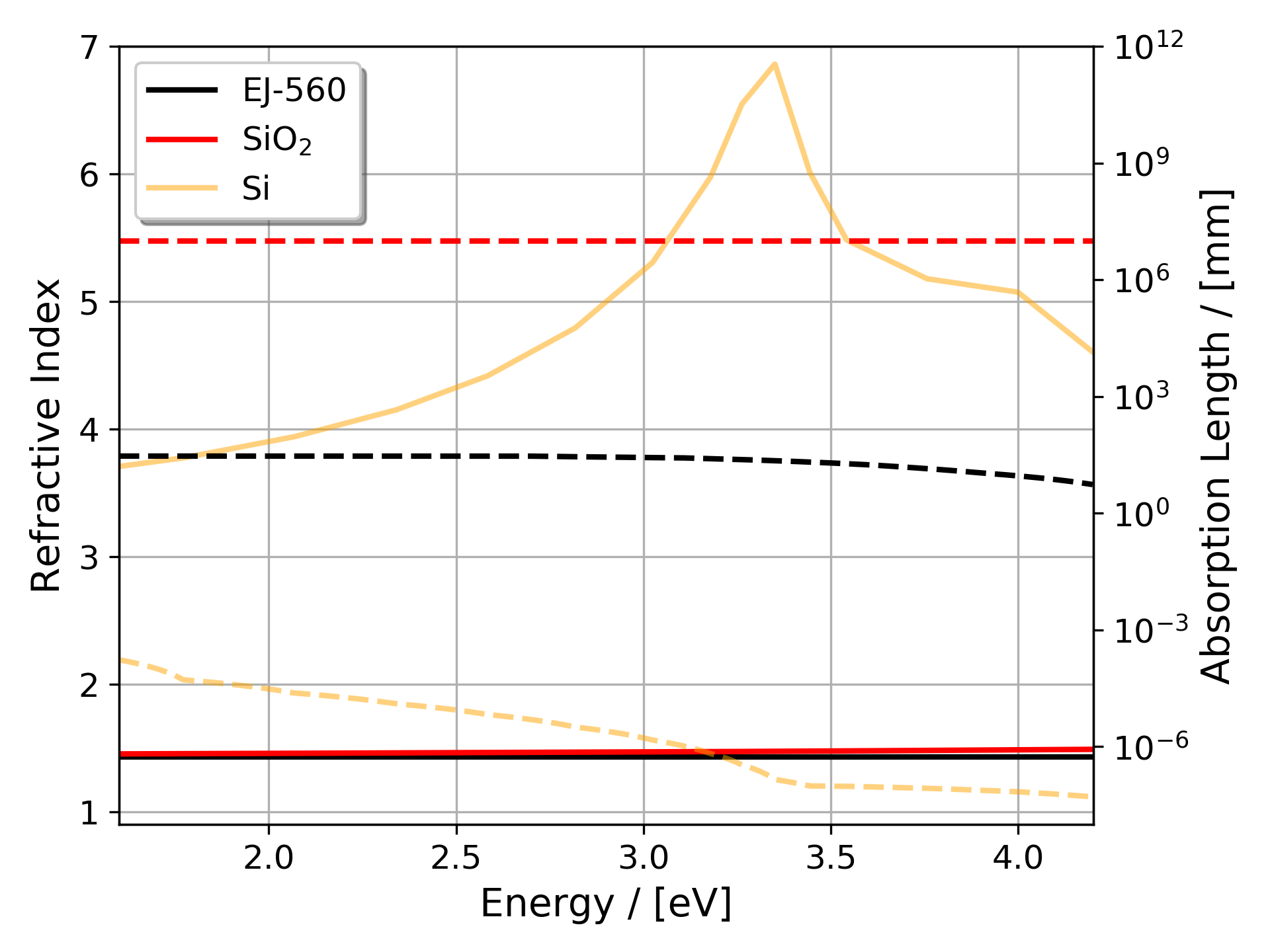}
    \end{subfigure}
    \begin{subfigure}
    \centering
        \includegraphics[width=0.45\textwidth]{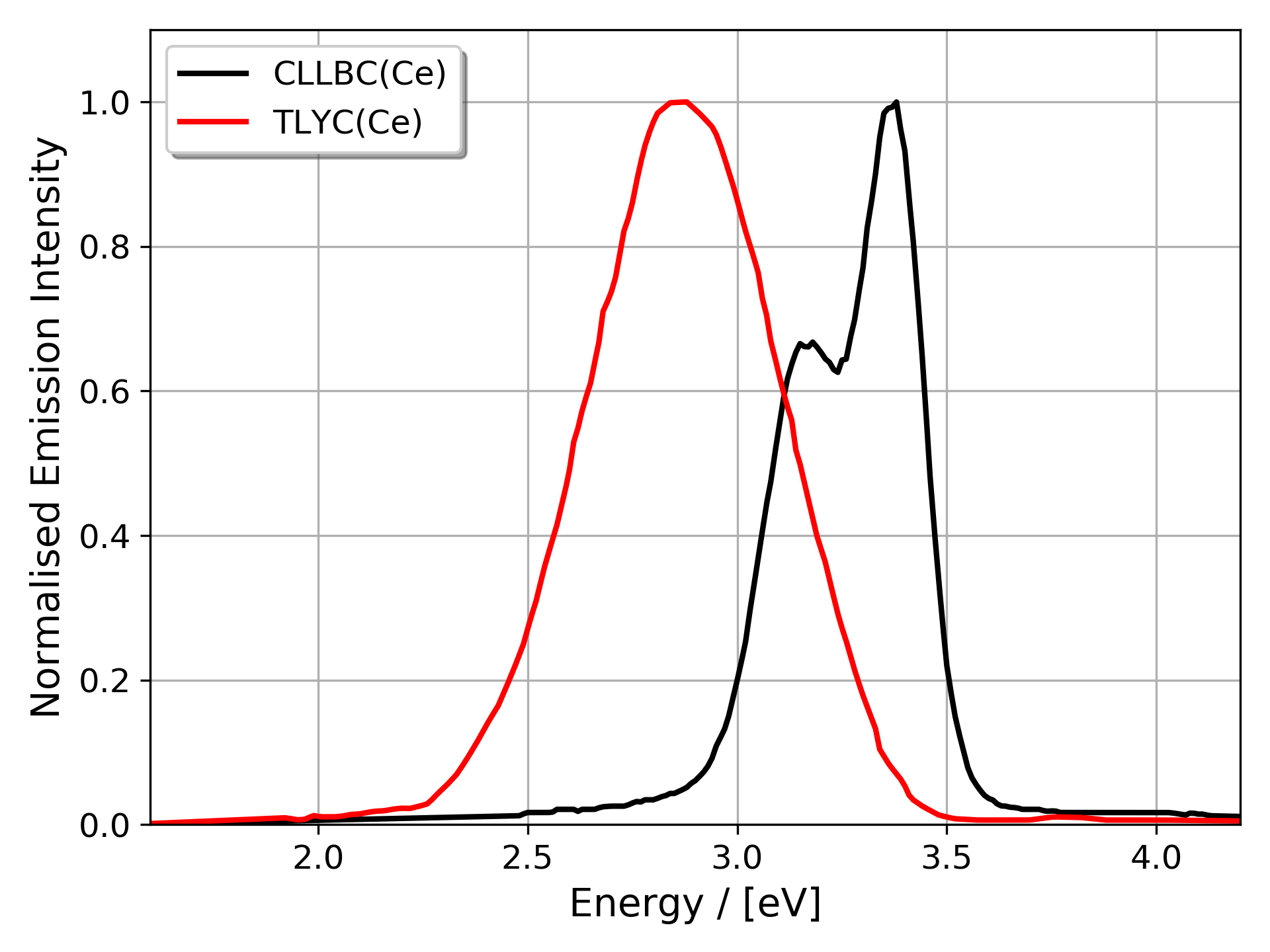}
    \end{subfigure}
\caption{Optical/scintillation properties of all materials utilised in the construction of the CLLBC(Ce) and TLYC(Ce) SiPM-based radiation detector model: [left] Optical pad (EJ-560), Onsemi SiPM glass (SiO$_2$), and Onsemi pixel (Si) material refractive index (solid line) and attenuation lengths (dashed line), and [right] CLLBC(Ce) and TLYC(Ce) crystal material normalised scintillation photon emission intensities (solid line).}
\label{fig:3}
\end{figure}

A detailed 1/2-inch CLLBC(Ce) and TLYC(Ce) SiPM-based radiation detector geometry model that includes all seven key components was implemented based on specifications from RMD. The detector was implemented as a 17.78 (D) $\times$ 17.78 (W) $\times$ 25.02 (H) mm aluminium housing with a 14.732 (D) $\times$ 14.732 (W) $\times$ 20.956 (H) mm dry air inner region. Within this region a 12.7 (D) $\times$ 12.7 (W) $\times$ 12.7 (H) mm CLLBC(Ce) and TLYC(Ce) crystal is made light tight on 5 sides through a 1.54 mm thick GORE diffuse reflector top cap and wrapped on 4 sides with 1.016 mm thick layer of Teflon tape. The bottom surface of the CLLBC(Ce) and TLYC(Ce) crystal is optically coupled with a 1 mm thick EJ-560 optical pad to an Onsemi ARRAYJ-60035-4P-BGA SiPM and 1.6 mm thick interface PCB. A final layer of 1.54 mm thick GORE diffuse reflector then fixes the components within the aluminium housing into place. Figure \ref{fig:3}, Table \ref{tab:1}, and Table \ref{tab:2} present the density, elemental composition, and optical/scintillation properties of all materials utilised in the construction of the CLLBC(Ce) and TLYC(Ce) SiPM-based radiation detector model. It should be noted that at present the optical attenuation length data for CLLBC(Ce) and TLYC(Ce) is unknown [Shirwadkar~U. \textit{Personal Communication}, February 2022]. However as both scintillator crystals are relatively small in volume and appear to be translucent in nature, this missing data is expected to have a minimal impact on this study. 

\subsection{Geant4 Application Physics and Optical Surface Modelling}

Gamma ray, electron, neutron, x-ray and alpha transport was simulated using the QGSP\_BIC\_HP\_EMZ physics constructor \cite{G4Phys2020,Arce2021} with atomic deexcitation, PIXE and radioactive decay enabled, a particle production length cut of 100 \textmu m, and a low-energy cut-off of 250 eV. Optical photon generation and transport was included for the processes of scintillation, absorption, refraction and reflection via the Geant4 implementation of the ``Unified" model \cite{G4Phys2020,Levin1996}. With the exception of the GORE diffuse reflector and Teflon to CLLBC(Ce) and TLYC(Ce) material interfaces (modelled as a dielectric-to-metal), all other material optical interfaces were modelled as dielectric-to-dielectric. Finally, every surface interface between two materials was described as a ground surface with surface roughness of 0.1 degrees as it is not possible for surfaces to be perfectly smooth \cite{VanderLaan2010,Nilsson2015,Brown2019,Brown2021}. 

\subsection{Geant4 Application Validation Simulations}

Five sets of simulations were undertaken with the developed Geant4 application to mimic the gamma ray and neutron irradiation experiments outlined above. For the three sets of gamma ray irradiation experiments with the planar Isotrack QSA Global sources (Co-60, Cs-137, and Am-241), each source’s physical geometry was implemented as a 51 (D) $\times$ 89 (w) $\times$ 6 (H) mm aluminium slab with a centralised recessed cylindrical foil 32 mm in diameter and 2.5 mm thick. At the centre of this cylindrical foil an infinitely thin 32 mm diameter plane was defined as the radiation emitting region. For the Spectrum Techniques encapsulated disk source (Eu-152) a 25.4 mm in diameter and 3.175 mm thick PMMA cylinder was implemented, and the radiation emission region was defined as a point location at its centre. To maximise computational efficiency, only the gamma ray emission was simulated for the radioactive decay of these four different radionuclides, with a total of 48 million decays simulated for each source and CLLBC(Ce) and TLYC(Ce) detector combination. All four radioactive sources were implemented with their centre aligned directly on top of the CLLBC(Ce) and TLYC(Ce) detector. 

For the Cf-252 radioactive source simulation, a 195 mm diameter/225 mm tall PMMA cylinder was orientated on top of the laboratory benchtop 100 mm away from simulated CLLBC(Ce) and TLYC(Ce) detector. The emission region of radiation from Cf-252 decay was approximated as a point source at the centre of the cylinder, with the process of radioactive decay modelled using Geant4’s G4RadioactiveDecay physics module \cite{G42003,G42006,G42016,G4Phys2020}. Finally, a total of 480 million decays were simulated for each CLLBC(Ce) and TLYC(Ce) detector combination. 

\section{Results and Discussion}

\begin{figure}[ht]    
    \centering
    \begin{subfigure}
    \centering 
        \includegraphics[width=0.425\textwidth]{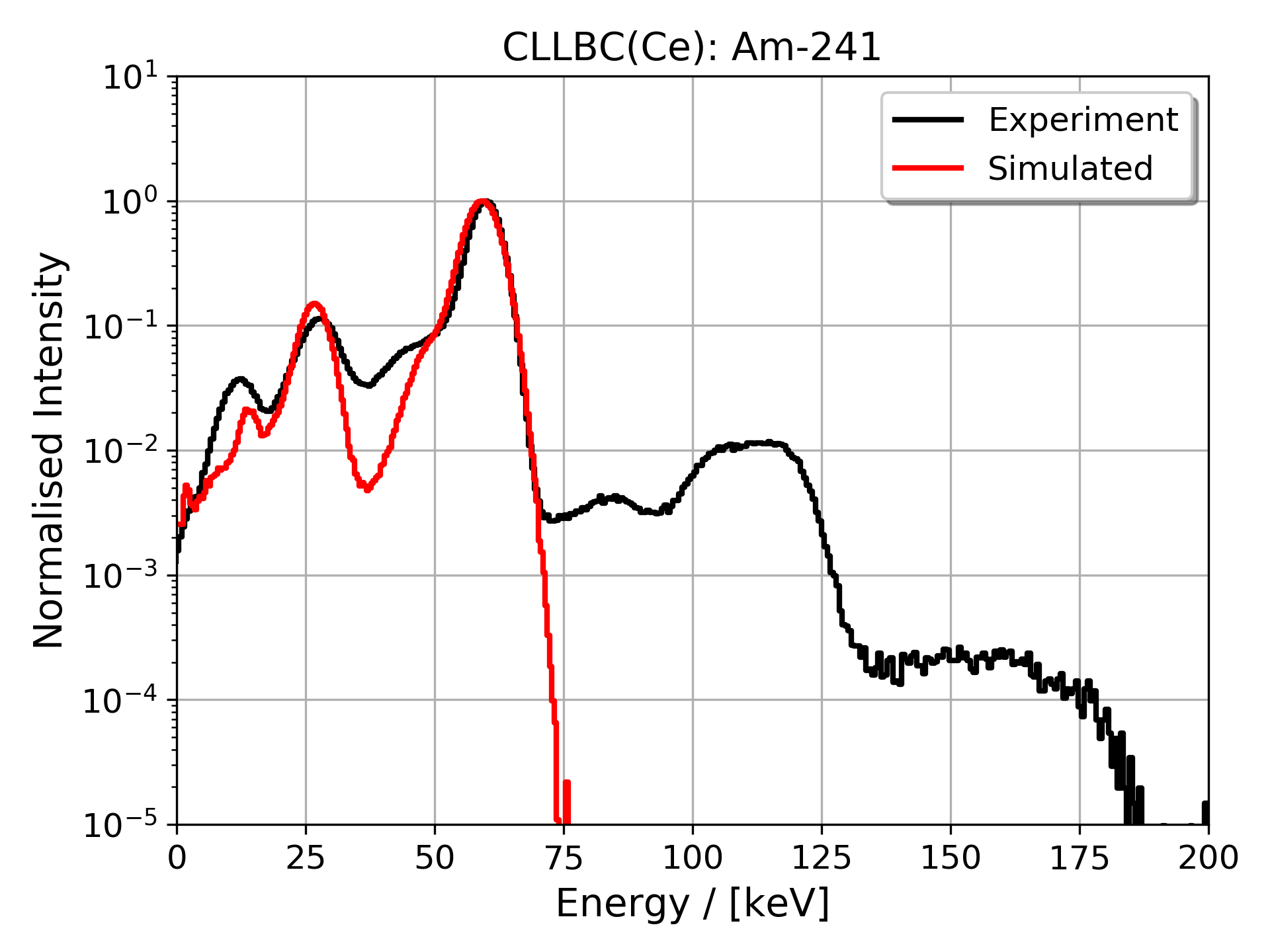}
    \end{subfigure}
    \begin{subfigure}
    \centering
        \includegraphics[width=0.425\textwidth]{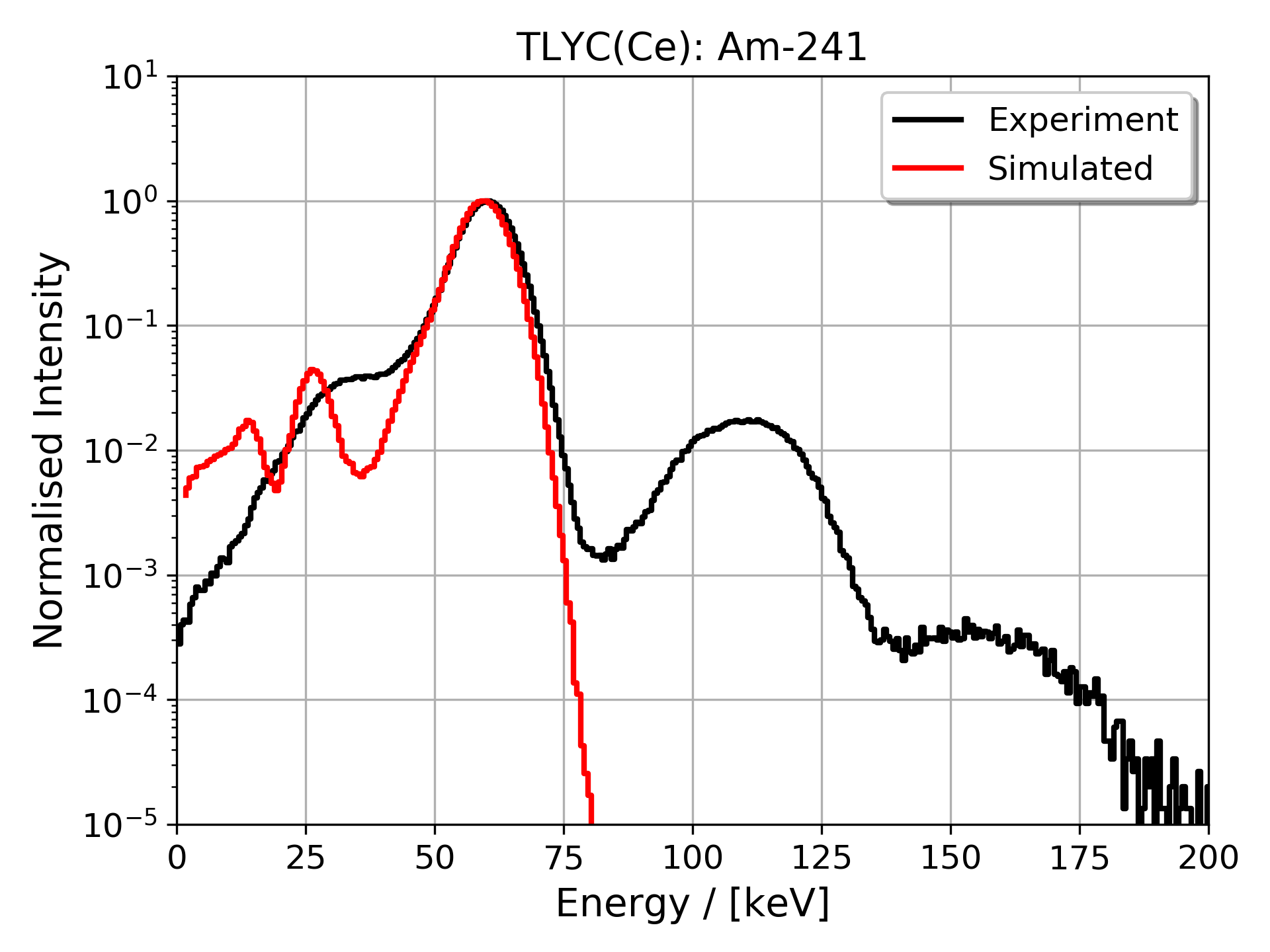}
    \end{subfigure}

    \begin{subfigure}
    \centering 
        \includegraphics[width=0.425\textwidth]{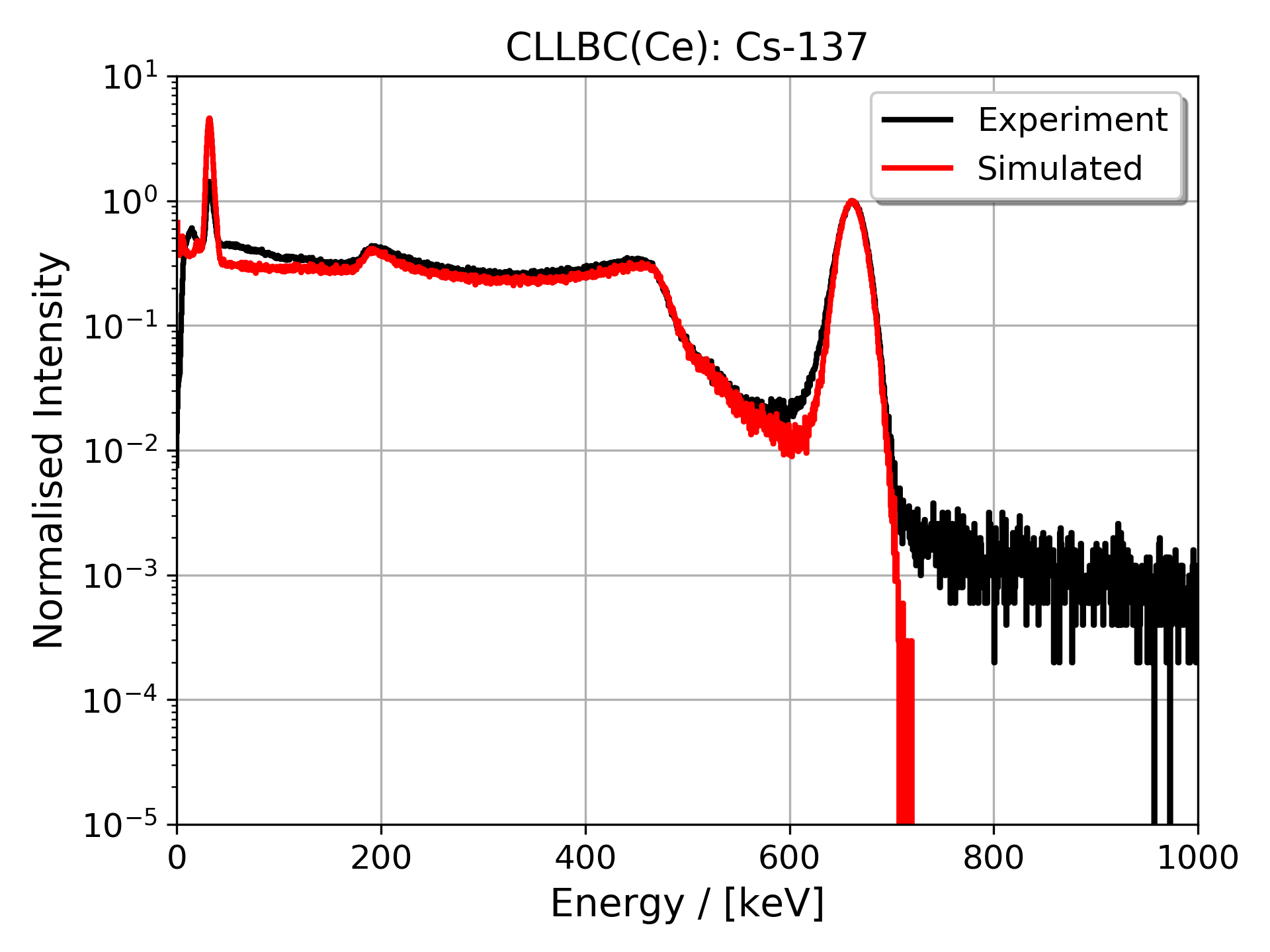}
    \end{subfigure}
    \begin{subfigure}
    \centering
        \includegraphics[width=0.425\textwidth]{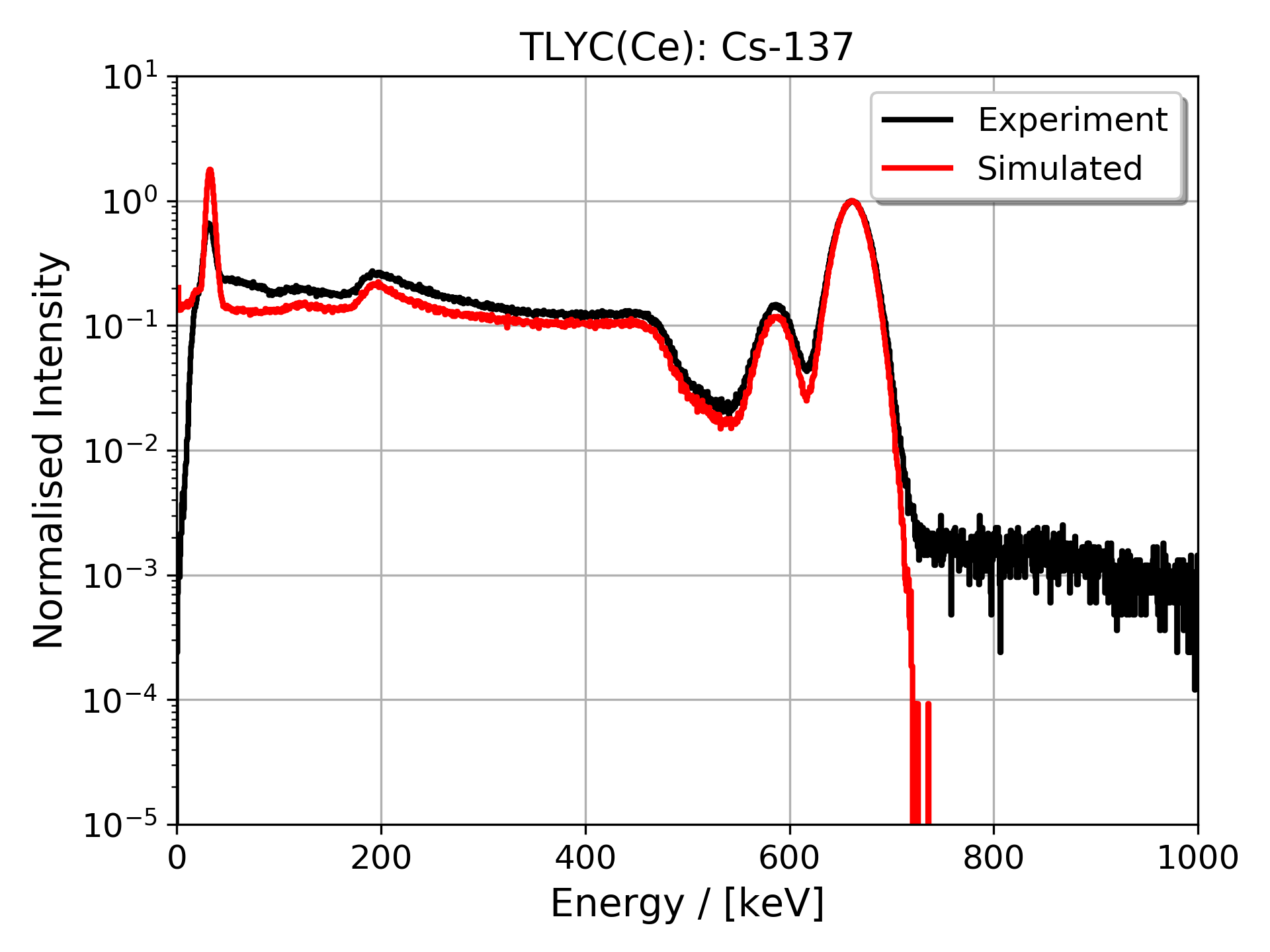}
    \end{subfigure}

    \begin{subfigure}
    \centering 
        \includegraphics[width=0.425\textwidth]{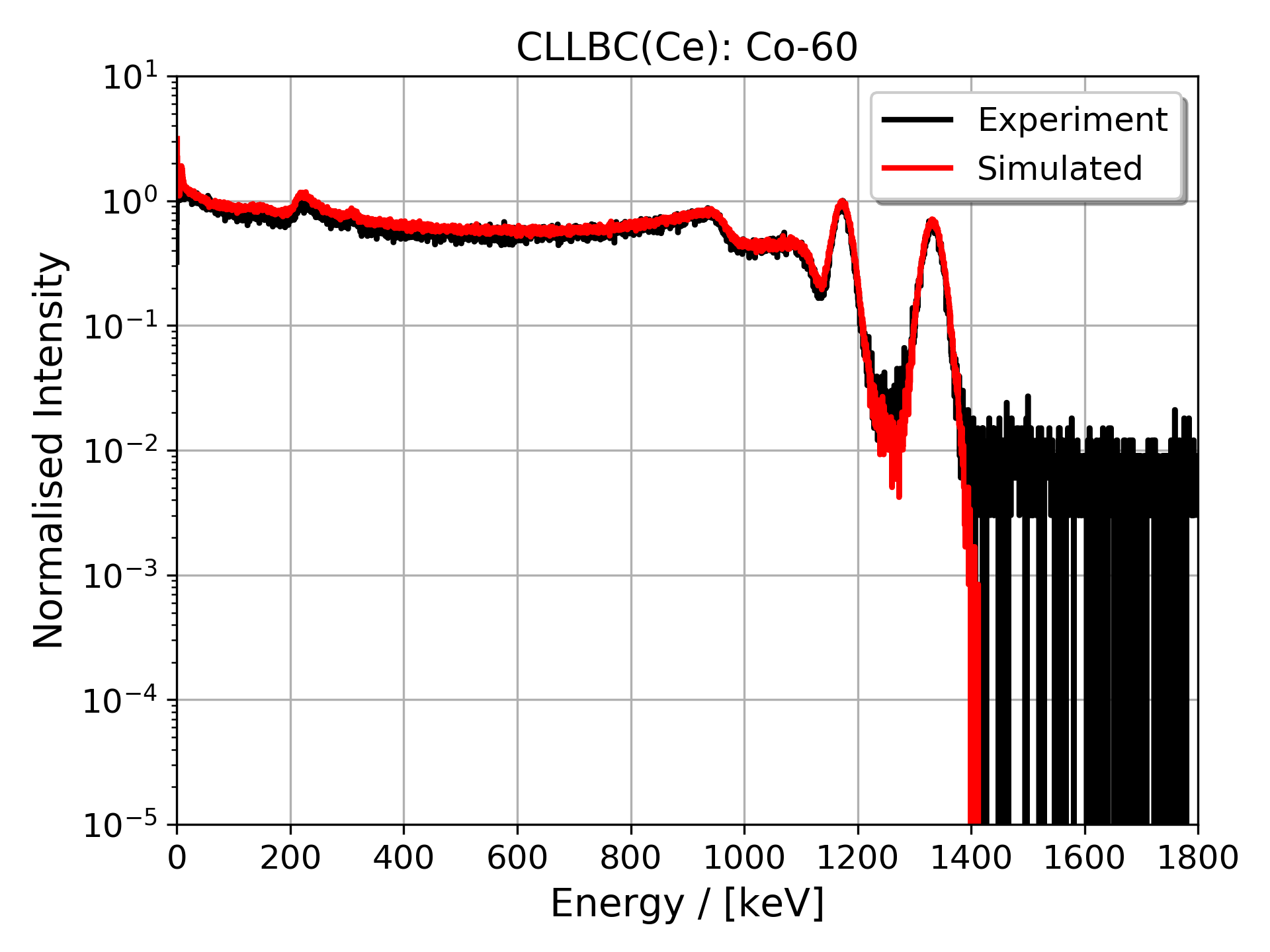}
    \end{subfigure}
    \begin{subfigure}
    \centering
        \includegraphics[width=0.425\textwidth]{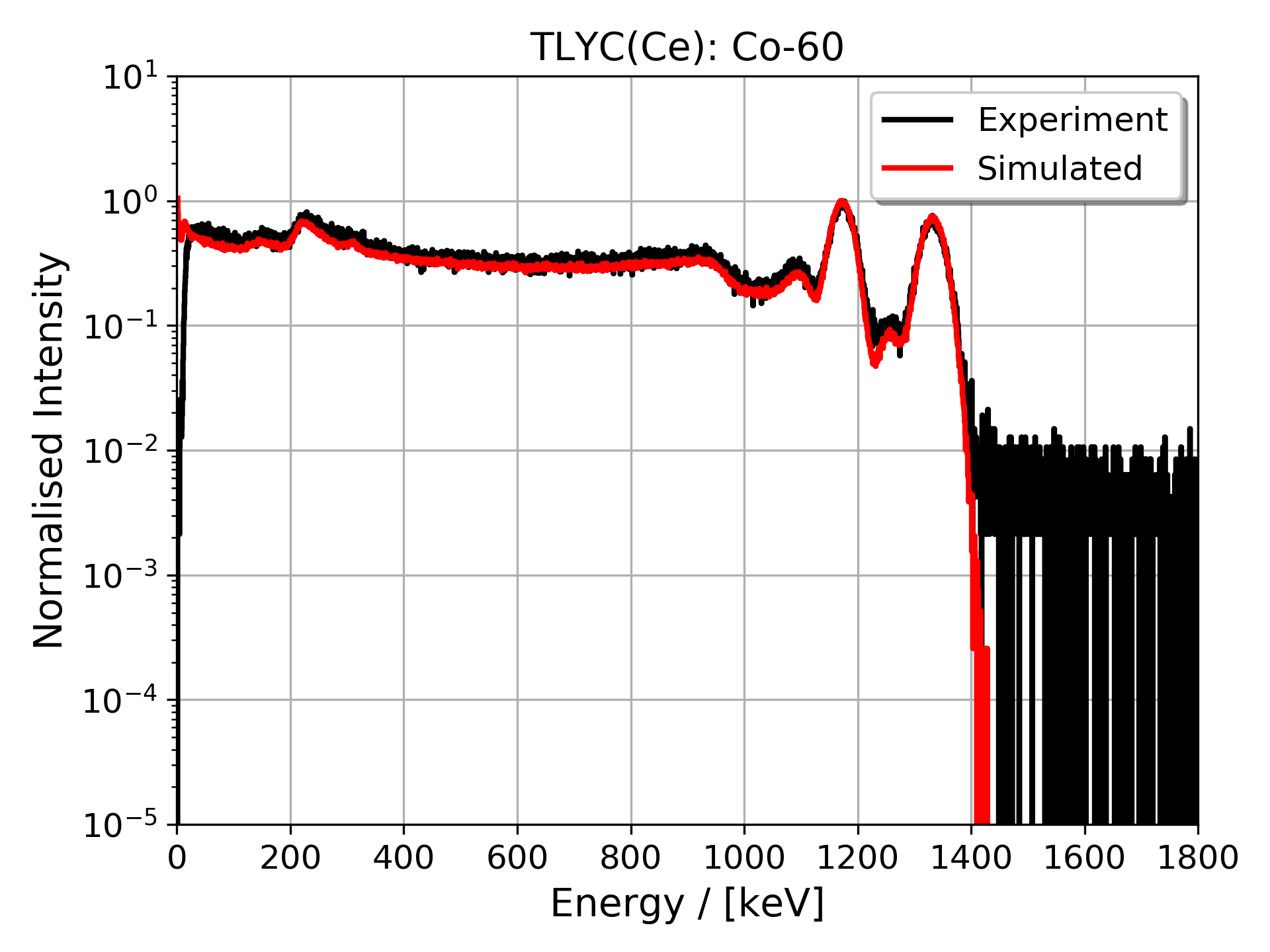}
    \end{subfigure}

    \begin{subfigure}
    \centering 
        \includegraphics[width=0.425\textwidth]{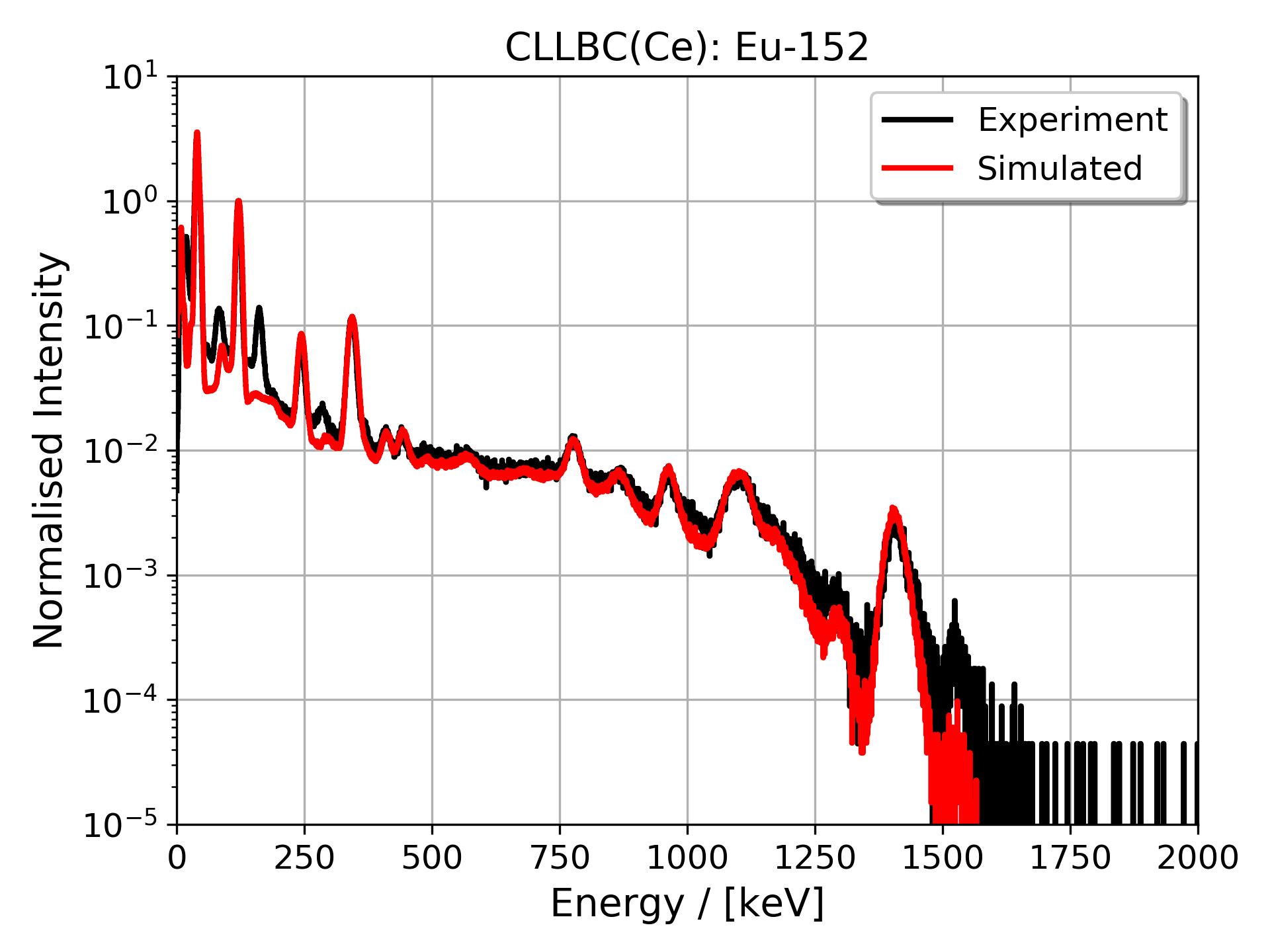}
    \end{subfigure}
    \begin{subfigure}
    \centering
        \includegraphics[width=0.425\textwidth]{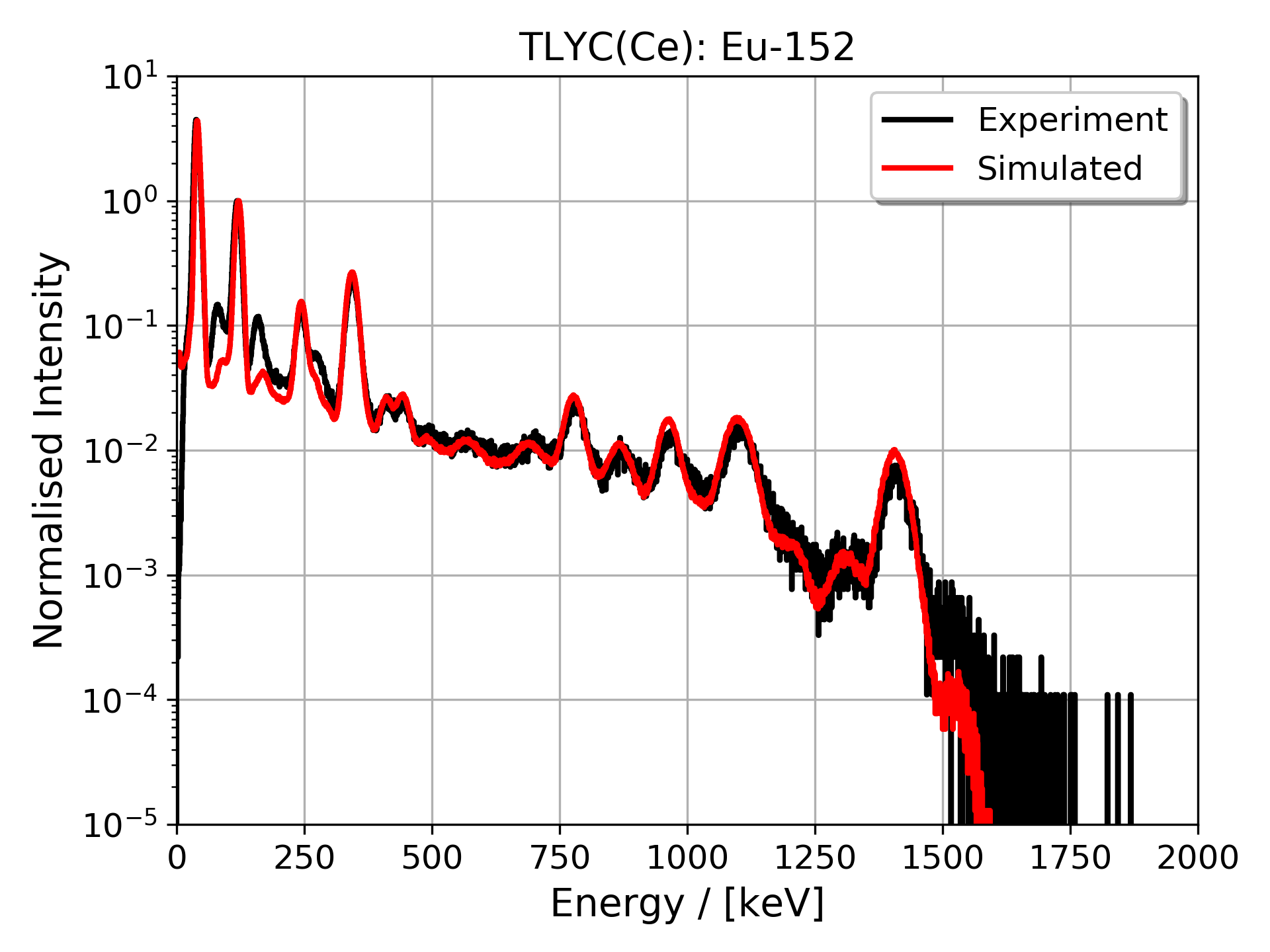}
    \end{subfigure}    
    
\caption{Experimental and simulated Geant4 gamma ray energy spectra from Am-241, Cs-137, Co-60, and Eu-152 sources obtained with the RMD 1/2-inch CLLBC(Ce) and TLYC(Ce) SiPM-based radiation detectors. Here the black and red line profiles represent the experimental and simulated data respectively. }
\label{fig:4}
\end{figure}

Figure \ref{fig:4} presents a comparison of the measured experimental and simulated gamma ray energy spectra of Am-241, Cs-137, Co-60, and Eu-152 sources obtained with the RMD 1/2-inch CLLBC(Ce) and TLYC(Ce) SiPM-based radiation detectors. Overall, a high level of correlation can be observed between major spectral features in the experimental and simulated energy spectra obtained with both radiation detectors (full energy gamma ray photopeaks, X-ray escape photopeaks, Compton edge, Compton backscatter peaks, and Compton plateau). In the specific case of the full energy gamma ray photopeaks for each predominate gamma ray emission of the four gamma ray sources, Table \ref{tab:3} shows that there is less than a 2\% and 1\% maximum difference in the Full Width at Half Maximum (FWHM) values of the CLLBC(Ce) and TLYC(Ce) energy spectra datasets respectively. Finally, the differences that can be observed between experimental and simulated data for all four gamma ray sources can be attributed to two primary factors: 1) approximations made in the modelling of the electronics response functions in the Geant4 application, and 2) multiple scattering of radiation within surrounding lab environment which was not included in the Geant4 application. 

Two clear examples of the shortcomings of the implemented electronics response modelling in the Geant4 application can be observed in both sets of Am-241 and Eu-152 energy spectra. In the Am-241 energy spectra a clear double 59.54 keV gamma ray double sum peak, and a double sum peak with background Compton scatter “continuum” can be observed in the right-hand side in the experimental energy spectra that is not present in their simulated counterparts. Additionally in the left-hand side of each Am-241 experimental energy spectra, the lack of a minimum pulse height threshold that mimics the behaviour of the XIA microDXP results in the simulated energy spectra significantly deviating from the experimental energy spectra below 20 keV for both radiation detectors. Whereas for the Eu-152 spectra, several additional “photopeaks” can be observed in the CLLBC(Ce) and TLYC(Ce) experimental energy spectra below 300 keV that are not present in the simulated data. These additional “photopeaks” in the experimental energy spectra are due to the coincidence summing of the characteristic k-shell X-rays with themselves, and the 121.78 keV and 244.70 keV gamma rays emitted during the radioactive decay of Eu-152. 

\begin{figure}[ht]    
    \centering
    \begin{subfigure}
    \centering 
        \includegraphics[width=0.425\textwidth]{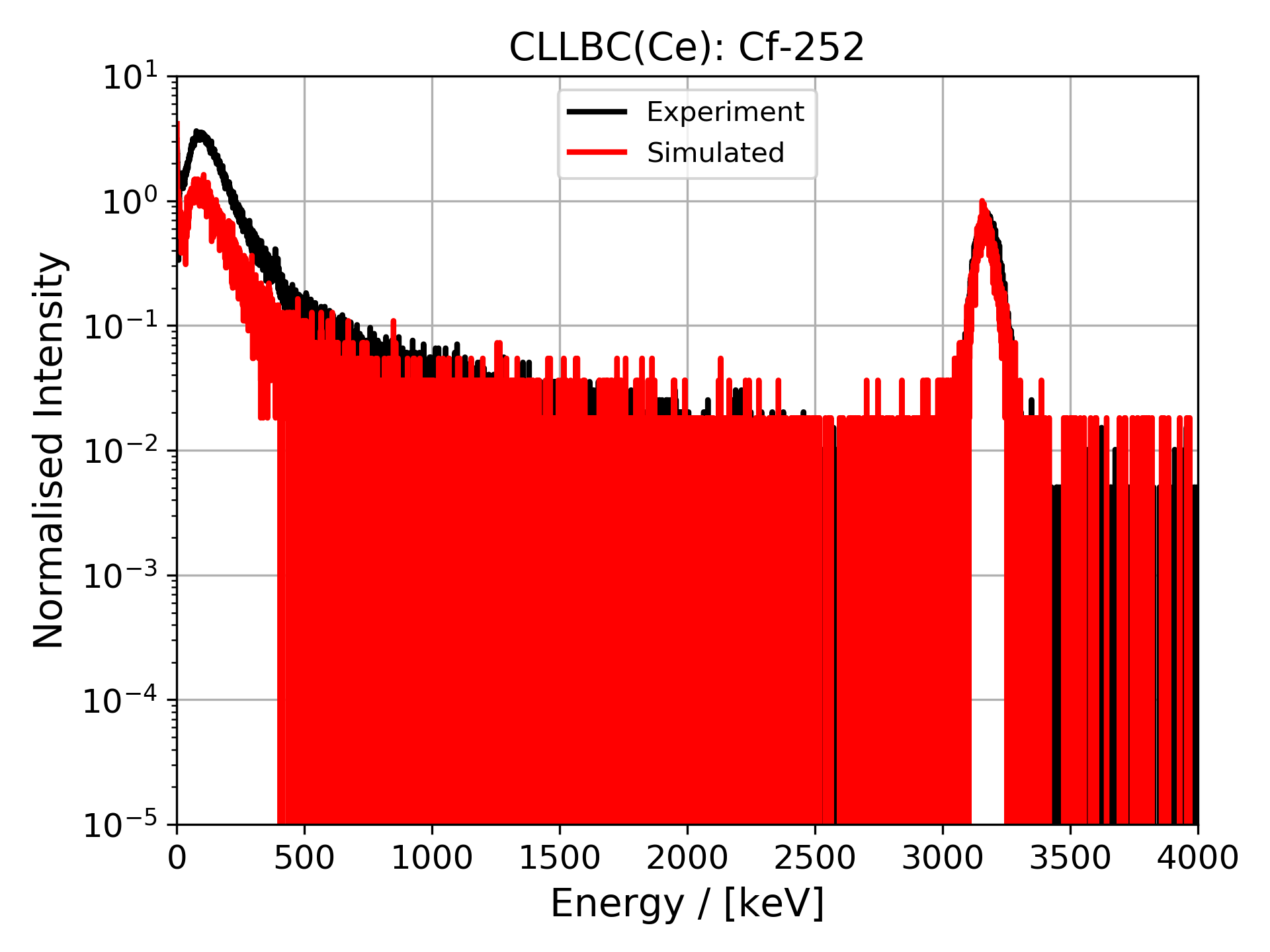}
    \end{subfigure}
    \begin{subfigure}
    \centering
        \includegraphics[width=0.425\textwidth]{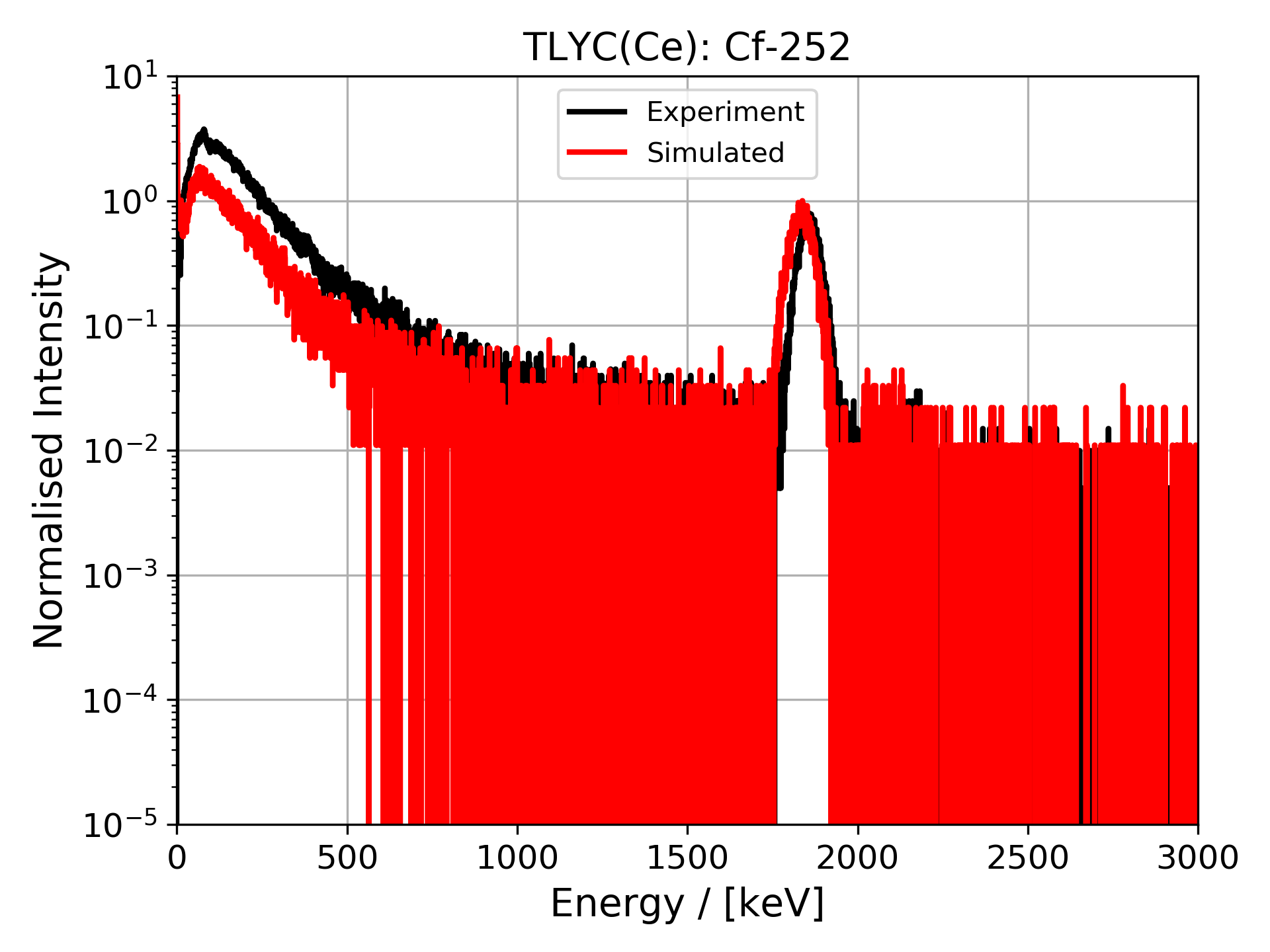}
    \end{subfigure}
    
\caption{Experimental and simulated Geant4 gamma ray energy spectra from a Cf-252 source obtained with the RMD 1/2-inch CLLBC(Ce) and TLYC(Ce) SiPM-based radiation detectors. Here the black and red line profiles represent the experimental and simulated data respectively. }
\label{fig:5}
\end{figure}

Figure \ref{fig:5} presents a comparison of the measured experimental and simulated mixed gamma ray/neutron spectra of Cf-252 obtained with the RMD 1/2-inch CLLBC(Ce) and TLYC(Ce) SiPM-based radiation detectors. For both sets of simulated energy spectra the Li-6 FEEP neutron capture peak spectral feature correlates to a high level of accuracy with their respective experimental energy spectra. Quantification of the centroid energy peak position and FHWM difference of each simulated Li-6 FEEP neutron capture peak shown in Table \ref{tab:4} illustrates that their is less than a 1\% and 2\% difference for the CLLBC(Ce) and TLYC(Ce) SiPM radiation detectors respectively. However, a significant difference between the simulated and experimental prompt gamma ray continuum relative to their respective Li-6 FEEP neutron capture peak can be observed in both sets of energy spectra. As with the gamma ray energy spectra results shown in Figure \ref{fig:4}, this observed difference in the prompt gamma ray continuum can be attributed to the multiple scattering of gamma rays within surrounding lab environment which was not included in the Geant4 application. 

The high level of correlation that can be observed between major spectral features in the energy spectra presented in Figures \ref{fig:4} and \ref{fig:5} illustrate that the optimised material composition, optical data tables, and physics transport settings implemented in the developed Geant4 application enable the realistic simulation of CLLBC(Ce) and TLYC(Ce) SiPM-based radiation detector response. Whilst there were observed deviations between the experimental and simulated energy spectra in the low energy region below each of the major spectral features, these differences can be primarily attributed to the impact of multiple scattering of radiation within surrounding lab environment which was not included in the Geant4 application. Additionally the assumptions that were made with regard to the impact of the lack of optical attenuation length data for CLLBC(Ce) and TLYC(Ce) were proven to be correct as indicated by the less than 2\% deviation between the experimental and simulated FWHM energy resolution for all FEEP/photopeaks observed. It is expected that this assumption would still hold true for simulations with CLLBC(Ce) and TLYC(Ce) detectors of different crystal geometries of similar scale (approximately 10 to 40 mm). However, above this physical limit determination of optical attenuation length would be needed in conjunction with further experimental testing and validation.

\section{Conclusion}

A Geant4 application was developed and material composition, optical data tables, and physics transport settings were implemented to enable the accurate simulation of the response of CLLBC(Ce) and TLYC(Ce) SiPM-based radiation detectors under both gamma ray and neutron irradiation. Experimental benchmarking for five different radioactive sources (Co-60, Cs-137, Eu-152, Am-241, and Cf-252) illustrated that this developed Geant4 application was able to reproduce the position and structure of all major spectral features (full energy gamma ray/neutron capture photopeaks, X-ray escape photopeaks, Compton edge, Compton backscatter peaks, and Compton plateau) to high level of accuracy. With these determined material compositions, optical data tables, and physics transport settings, it will be possible to freely explore different CLLBC(Ce) and TLYC(Ce) detector designs and test their deployment in-silico for a variety of scenarios relevant to the fields of homeland and nuclear safety/security.

\section*{Acknowledgments}

J.~M.~C.~Brown would like to acknowledge Dr. Urmila Shirwadkar and Mr. Joshua Tower of Radiation Monitoring Devices Inc. (U.S.A.) for their helpful clarification of construction and material properties of the supplied 1/2-inch CLLBC(Ce) and TLYC(Ce) SiPM-based radiation detectors. This work was funded under Project Arrangement 10256 in collaboration with the Defence Science and Technology Group of the Australian Department of Defence (Australia).

\newpage

\newpage

\newpage

\newpage

\newpage

\newpage

\begin{sidewaystable}
\centering
\begin{tabular}{||c|c|c|c|c|c||}
\hline
\hline
             &              &           &            & Optical        &          \\
 Material    & Density      & Elemental & Refractive & Reflectivity / & Reference \\
             & (g/cm$^{3}$) & Composition & Index      & Absorption   & \\
 \hline
 \hline             
  Dry Air    & 1.29$\times$10$^-3$ & C (0.01\%), N (75.52\%), & 1 & - & Geant4 Material  \\
             &       &  O (23.19\%), Ar (1.28\%) &  &  &   Database \cite{G42016,G4Phys2020} \\
 \hline   
  PCB        & 1.86  & SiO$_2$ (52.8\%), H$_1$C$_1$O$_1$ (47.2\%) & - & 0\% / 100\% & \cite{Brown2019,Brown2021} \\
 \hline   
  SiPM Glass & 2.203 & SiO$_2$ & See Fig. \ref{fig:3} & See Fig. \ref{fig:3} & \cite{Brown2019,Brown2021} \\
 \hline   
  SiPM Pixel &  2.33 & Si & See Fig. \ref{fig:3} & See Fig. \ref{fig:3} & \cite{Philipp1960} \\
 \hline                
  EJ-560 Optical Pad  &  1.03  & H$_6$C$_2$O$_1$Si$_1$ & See Fig. \ref{fig:3} & See Fig. \ref{fig:3}  & \cite{Eljen2021} \\  
 \hline                
  GORE Diffuse Reflector  &  0.65  & C$_2$F$_4$ & - & 96\% / 4 \%  & \cite{Janecek2012} \\ 
 \hline
  Teflon Tape  &  2.2  & C$_2$F$_4$ & - & 88\% / 12 \%  & \cite{Janecek2012} \\ 
 \hline   
 \hline
\end{tabular}
\\
\caption[]{Density, elemental composition, and optical properties of non-scintillator RMD detector materials.}
\label{tab:1}
\end{sidewaystable}

\newpage

\begin{sidewaystable}
\centering
\begin{tabular}{||c|c|c|c|c|c|c|c||}
\hline
\hline
            &             &                &             &                     &   Optical Decay   &  Resolution Scale                &           \\
Material      & Density      & Elemental   & Refractive  & Optical Yield,      &  Time Constants   & (at 662 keV),                    & Reference \\
              & (g/cm$^{3}$) & Composition & Index       & Emission Spectrum   &       (ns)        & Birks' Constant                  & \\
              &              &             &             &                     &                   & (mm/keV)                         & \\
 \hline 
 \hline    
CLLBC(Ce)  & 4.06       & Cs$_{2}$Li$_{1}$La$_{1}$Br$_{4.8}$Cl$_{1.2}$ &  1.9  & 50 Photons per keV, & Fast: 130 (82.5\%) & 2.167,  & \cite{Shirwadkar2012,Hawrami2016a} \\
           &        & (2\% Ce doping)            &                          & See Fig. \ref{fig:3}  & Slow: 784 (17.5\%) & 3.85  &  \\ 
              &              &  (95\% Li-6 enrichment)           &             &                          &                   &             & \\           
 \hline    
TLYC(Ce)   & 4.5       & Tl$_2$Li$_1$Y$_1$Cl$_{6}$ &  2.4 & 29 Photons per keV, & Fast: 71 (32.3\%) & 2.345, & \cite{Hawrami2015,Hawrami2016b} \\
           &        & (3\% Ce doping)           &                          & See Fig. \ref{fig:3} & Slow: 537 (67.7\%) &  14.2 & \\  
              &              &  (95\% Li-6 enrichment)           &             &                          &                   &             & \\    
 \hline   
 \hline
\end{tabular} 
 \caption[]{Density, elemental composition, and optical properties of CLLBC(Ce) and TLYC(Ce).}
\label{tab:2}
\end{sidewaystable}

\newpage

\begin{table}
 \centering
\begin{tabular}{||c|c|c|c|c||}

\hline
\hline
Radioactive Source / & \multicolumn{2}{c|}{CLLBC(Ce)} & \multicolumn{2}{c||}{TLYC(Ce)} \\
Gamma Ray [keV]      & \multicolumn{2}{c|}{FWHM[\%]} & \multicolumn{2}{c||}{FWHM[\%]} \\
\hline
                     & Experimental       &   Simulated            & Experimental      &   Simulated            \\
 \hline
 \hline             
Am-241: 59.54        & 10.28      & 12.43             & 17.29      & 17.02              \\
Eu-152: 121.78       & 7.17       & 8.60              & 11.72      & 11.56              \\
Eu-152: 244.70       & 5.15       & 6.18              & 7.47       & 7.90               \\
Eu-152: 344.28       & 4.41       & 5.25              & 6.87       & 7.00               \\
Cs-137: 661.66       & 3.76       & 3.90              & 5.08       & 5.01               \\
Eu-152: 778.91       & 3.32       & 3.53              & 4.53       & 4.67               \\
Eu-152: 964.06       & 2.40       & 3.16              & 3.71       & 4.08               \\
Co-60: 1173.23       & 2.42       & 2.82              & 3.60       & 3.70               \\
Co-60: 1332.49       & 2.35       & 2.78              & 3.44       & 3.57               \\
 \hline   
 \hline
\end{tabular}
\\
\caption[]{Comparison of experimental and simulated RMD 1/2-inch CLLBC(Ce) and TLYC(Ce) SiPM-based radiation detector full energy gamma ray photopeak resolutions (Full Width at Half Maximum [FWHM]) from the energy spectra presented in Figure \ref{fig:4}.}
\label{tab:3}
\end{table}

\newpage

\begin{table}
 \centering
\begin{tabular}{||c|c|c|c|c||}

\hline
\hline
Material & \multicolumn{2}{c|}{Experimental} & \multicolumn{2}{c||}{Simulated} \\

\hline
                     & Centroid [keV]       &   FWHM[\%]            & Centroid [keV]          &   FWHM[\%]             \\
 \hline
 \hline             
CLLBC(Ce)       & 3172.73       & 2.49             & 3164.93      & 2.33              \\
TLYC(Ce)        & 1857.41       & 3.38             & 1835.25      & 4.11              \\
 \hline   
 \hline
\end{tabular}
\\
\caption[]{Comparison of experimental and simulated RMD 1/2-inch CLLBC(Ce) and TLYC(Ce) SiPM-based radiation detector centroid position and FHWM for the Full Energy Equivalent Photopeak (FEEP) generated via Li-6 neutron capture presented in Figure \ref{fig:5}.}
\label{tab:4}
\end{table}

\end{document}